\documentclass[twocolumn,lineno,dvipsnames, usenames]{aastex62}
\usepackage{amssymb,amsmath,verbatim,mathtools,needspace,enumitem,etoolbox,graphicx,physics,microtype,ulem}
\usepackage{xspace}
\definecolor{linkcolor}{rgb}{0.0,0.3,0.5}
\definecolor{urlcolor}{rgb}{0.27,0.55,0.}
\definecolor{funcolor}{rgb}{0.65, 0.16, 0.16}
\usepackage[all]{hypcap}
\usepackage{slashbox,booktabs,amsmath}

\def\apjl{\rm{ApJ}}
\def\apjs{\rm{ApJS}}
\def\aap{\rm{A\&A}}

\newcommand{\mc}{\ensuremath{\mathcal{M}}\xspace}

\newcommand{\chieff}{\ensuremath{\chi_{\mathrm{eff}}}\xspace}

\newcommand{\Msun}{M_{\odot}}
\newcommand{\msun}{\ensuremath{M_{\odot}}\xspace}

\newcommand{\beqn}{\begin{eqnarray}}
\newcommand{\enqn}{\end{eqnarray}}

\newcommand{\LIGOlabMIT}{\affiliation{LIGO Laboratory, Massachusetts Institute of Technology, 185 Albany St, Cambridge, MA 02139, USA}}
\newcommand{\MKI}{\affiliation{Department of Physics and Kavli Institute for Astrophysics and Space Research, Massachusetts Institute of Technology, 77 Massachusetts Ave, Cambridge, MA 02139, USA}}

\begin{document}

\title{The reliability of the low-latency estimation of binary neutron star chirp mass}

\author{Sylvia Biscoveanu} \LIGOlabMIT \MKI
\email{sbisco@mit.edu}
\author{Salvatore Vitale} \LIGOlabMIT \MKI
\email{salvatore.vitale@ligo.org}
\author{Carl-Johan Haster} \LIGOlabMIT \MKI

\date{\today}

\begin{abstract}
The LIGO and Virgo Collaborations currently conduct searches for gravitational waves from compact binary coalescences in real-time.
For promising
candidate events, a sky map and distance estimation are released in low-latency, to facilitate their electromagnetic follow-up. 
Currently, no information is released about the masses of the compact objects. 
Recently, \cite{M5} have suggested that knowledge of the chirp mass of the detected binary neutron stars could be useful to prioritize the electromagnetic follow-up effort, and have urged the LIGO-Virgo collaboration to release chirp mass information in low-latency. 
One might worry that low-latency searches for compact binaries make simplifying assumptions that could introduce biases in the mass parameters: neutron stars are treated as point particles with dimensionless spins below $0.05$ and perfectly aligned with the orbital angular momentum. 
Furthermore, the template bank used to search for them has a finite resolution. In this paper we show that none of these limitations can introduce chirp mass biases larger than $\sim 10^{-3}~M_\odot$. Even the total mass is usually accurately estimated, with biases smaller than 6\%. The mass ratio and effective inspiral spins, on the other hand, can suffer from more severe biases.

\end{abstract}

\section{Introduction}

The discovery of the binary neutron star (BNS) merger GW170817~\citep{gw170817} in both the gravitational-wave (GW) and electromagnetic (EM) bands~\citep{gw170817_multimessenger} has shown the great potential of multi-messenger astrophysics. 
Besides proving that  BNS mergers are the progenitors of at least some short gamma-ray bursts (sGRBs)~(e.g. \cite{ eichler_bns_progenitors, gw170817_grb, grb170817A_fermi, grb170817A_integral}), that event has given a glimpse at the energetics of BNS mergers at all frequencies and at the formation of a kilonova~(e.g. \cite{kilonova_Li_1998, gw170817_kilonova1, gw170817_kilonova2, gw170817_kilonova3, gw170817_kilonova4, gw170817_kilonova5}) and allowed for a standard-siren measurement of the Hubble constant~\citep{hubble_constant}.
In the hours and weeks following the detection of the GWs and the sGRB, virtually all telescopes in the world were observing that part of the sky.

As the number of interesting GW sources to follow-up in the EM band increases, it might become necessary to prioritize and decide which ones are especially worth following up. 
During their third observing run,  which began in April 2019 and is scheduled to last one year, the GW detectors LIGO~\citep{aLIGO, Harry_2010} and Virgo~\citep{aVirgo} might detect up to 10 BNS systems, a number that will become even higher as further upgrades are made to the detectors~\citep{LVCEMGuide}. 

The LIGO-Virgo Collaborations (LVC) search for compact binary coalescences, including BNSs, in real time~\citep{gwtc1, mbta, pycbc_usman,spiir, gstlal_messick,pycbc_nitz,gstlal_sachdev}. For candidate events below some pre-determined false-alarm threshold, a public alert is released in low-latency, together with a skymap and a distance estimation, to enable their EM follow-up~\citep{LVCEMGuide}. 
At the time of writing, no estimation of the component masses (or derived quantities, such as the chirp mass or total mass) is released. 
Instead, the alert contains the probability that the source belongs to
a specific category (BNS, binary black hole, neutron star--black hole, mass
gap, and terrestrial)~\citep{LVCEMGuide} and the probability that it will emit light, according to published models~\citep{p_has_remnant, p_astro_calc}.

Recently, \cite{M5} have argued that releasing an estimate of the chirp mass (defined below)
of BNSs in low latency will allow EM astronomers to better allocate follow-up resources, based on which gravitational-wave events have the largest potential for enriching our understanding of merger outcomes and providing equation-of-state
constraints with a multi-messenger observation. 
Kilonova modeling for GW170817 revealed that most of the ejecta came from accretion disk outflows in the post-merger phase rather than from dynamical processes during the merger~\citep{Metzger_2008, Fernandez_and_Metzger, Perego, Siegel, Fernandez}. 
The accretion disk properties depend most sensitively on the total mass of the binary, rather than the mass ratio, which is not well-constrained by GW observations~\citep{Coughlin, Radice}. 
\cite{M5} then argue that the total mass can be inferred from the chirp mass of the binary given a prior on the mass ratio informed by galactic neutron star binaries. 
The total mass can then be used to predict the properties of the accompanying EM signal, allowing observers to prioritize binaries with the largest potential for placing constraints on the neutron star equation of state according to their ``multi-messenger matrix".

In this paper we investigate whether the estimation of BNS mass parameters that can be obtained in low-latency by the search algorithms is accurate. 
There are several reasons why it might not be.
Compact binary coalescence (CBC) search algorithms are currently based on matched filtering (see e.g.~\cite{findchirp, time_domain_mf}). 
While the details of its implementation depend on the algorithm and involve subtle points (see Sec.~\ref{Sec.Discussion} for a discussion of some of them), the concept is quite simple: a bank of waveforms is pre-built and used to filter the data~\citep{template_bank1, template_bank2, template_bank3, template_bank4}. 
The waveform template that yields the highest signal-to-noise (SNR) ratio, or an equivalent derived quantity (e.g. a chi-square weighted SNR, or a likelihood ratio~\citep{Cannon_likelihood_ratio})~\citep{mbta, pycbc_usman, spiir, gstlal_messick, pycbc_nitz,  gstlal_sachdev}, is selected and provides a point-estimate of some of the source's intrinsic parameters: masses and spins. 
We stress that detection algorithms are sensitive to the detector-frame mass, which is larger than the astrophysically relevant source-frame mass by a factor of $(1+z)$, with $z$ the redshift of the source. 
Unless otherwise indicated, we will only deal with the detector-frame chirp (or total) mass in this paper, and go back to the difference between these two quantities in Sec.~\ref{Sec.Discussion}.
Size and placement (i.e. the values of the intrinsic parameters of the templates) of the bank are chosen to guarantee a minimal overlap between expected GW signals and at least one waveform in the bank. 
Usually, one requires a minimal overlap of 97\% percent~\footnote{As the availability of computational resources improves, the minimum match is increasing, particularly in the binary black hole part of the bank where detections are more frequent, and increasing the minimum match only adds a small number of templates compared to the overall size of the bank~\citep{pycbc_template_bank}.}, where the overlap is defined as the inner product between the source waveform and the template waveform, weighted by the instrument noise spectral density and maximized over the extrinsic parameters (sky position, distance, etc.)~\citep{gstlal_template_bank}. 

However, recovering most of the SNR of the source signal does not necessarily imply that the point-estimate of its masses and spins is accurate. 
While this possible inaccuracy is especially relevant for high-mass systems, such as binary black holes, due to the smaller numbers of templates in that region~\citep{gw150914_searches, gwtc1}, it might  also be an issue for BNSs.
In addition, biases might arise due to missing physics in the template bank.
To keep the size of the template bank manageable, the spins of the compact objects are assumed to be perfectly aligned with the orbital angular momentum~\footnote{Allowing for generic spin tilt angles would double the dimensionality of the template bank, from 4 to 8 parameters, and result in a loss of sensitivity for sources with aligned spins~\citep{2016PhRvD..94b4012H}.}, and at least in the part of the bank that targets BNSs, the magnitude of the dimensionless spin is limited to $\leq 0.05$.~\citep{alignd_spin_bbh1, aligned_spin_bbh2}.
Furthermore, corrections to the point-particle approximation, such as the presence of tidal effects in neutron stars, are entirely neglected. 

All of these factors might introduce biases in the point-estimate of the mass and aligned-spin parameters obtained from the template bank in low-latency.
A last reason  is that the region of the template bank used to search for BNSs uses a very simple waveform model (\texttt{TaylorF2}~\citep{Buonanno:2009zt}) which does not include any post-inspiral GW signal. While the choice of waveform model should have a minimal effect on the recovery of BNS parameters, since these systems merge at high frequencies where the sensitivity of the detector is poor, this needs to be explicitly checked.

It is worth stressing that these limitations can be lifted when high-latency dedicated source characterization algorithms are run on candidate events.
These are usually based on stochastic sampling across a continuous parameter space and provide posterior distributions on the source parameters~\citep{lalinference,bilby}, while also being able to use more sophisticated waveform families that can account for larger spins with arbitrary orientation, tidal effects, and post-inspiral signal.
The extra complexity is compensated by a longer run time, usually hours or days depending on the mass of the source (lower mass systems take longer to analyze due to their longer in-band duration).
Alternative approaches also exist if only the intrinsic parameters (detector-frame masses, spins, tidal deformability) are of interest and a measurement of the extrinsic parameters is not necessary \citep{RIFT}.

In this paper we show that the estimation of the (detector frame) chirp mass of BNSs obtained by low-latency search algorithms is a) not significantly biased even when the true signal has features which are not included in the template bank, and b) usually consistent with what would be obtained by more sophisticated follow-up parameter-estimation codes, in spite of the missing physics and the finite size of the template bank. 
For a GW170817-like source, the difference between the source-frame and the detector-frame chirp mass would be much larger than the biases we find, making that the dominant error if the chirp mass measured by the search is interpreted directly as the astrophysical chirp mass.
Depending on the level of systematic uncertainty that can be tolerated, the total mass could also be used, as we find biases smaller than $6\%$ in all cases. 
Conversely, the mass ratio can suffer from large offsets, although these are not much larger than the statistical uncertainty obtained from analyses using more sophisticated higher-latency source characterization codes.

Our study suggests that the low-latency chirp mass point estimate for BNSs produced by current algorithms is adequate to inform EM follow-up strategies.

\section{Method and results}

\subsection{Biases from the template bank}\label{sec.ResultBanksim}

We first create a set of simulated BNS signals with features that are not accounted for in the template bank: large spin magnitude, spin misalignment and possibly tidal effects.
To isolate the contribution of various terms, we proceed by gradually increasing the complexity of the simulated signals.

To verify if and how much the results depend on the region of the parameter space that is being probed, we consider two different values of the mass ratio 

$$q\equiv m_2/m_1\;\mathrm{ with }\; m_2\leq m_1$$
and three values of chirp mass 

$$\mc\equiv \frac{(m_1 m_2)^{3/5}}{(m_1+m_2)^{1/5}}$$
where $m_1$ and $m_2$ are the component masses. Specifically, the signals we simulate have all possible combinations of

$$q \in [0.8,1.0]; \quad \mc \in [1.0, 1.15, 1.35]~\Msun.$$
These ranges are consistent with what is used in~\cite{M5}.

For each value of mass ratio and chirp mass, we create five BNS signals that differ by the value of the NS spin tilt angle, i.e. the angle between the spins and the orbital angular momentum, from $0^\circ$ (spins aligned with the orbital angular momentum vector) to $90^\circ$ (spins in the orbital plane). 
The inclination angle, i.e. the angle between the orbital angular momentum and the line of sight, is $35^\circ$ for all sources, which is near the peak of the distribution of orientations for detectable BNS systems~\citep{Schutz_2011, Chen_2018}.

For the first set of simulations, we assign to all the sources a spin magnitude $a=0.05$, equal to the maximum allowed by the template bank, and we do not include tidal effects. 
We use the \texttt{IMRPhenomPv2\_NRTidal} waveform~\citep{IMRPhenomPv2, nr_tidal2, nr_tidal1}. 
The distance of the sources is chosen such that they have an optimal {network} SNR of 35~\citep{GW150914PE} (in a network made of the two LIGO detectors and Virgo, all at design sensitivity~\citep{aLIGO, Harry_2010, aVirgo}, comparable to GW170817. 
We focus on loud events as for these the effects of potential systematic uncertainties and/or biases should be more prominent. {This is because the bias is an inherent property of the template bank for a given set of intrinsic parameters, and the statistical error shrinks with increasing SNR, so any bias will be more statistically significant for louder events.}

The simulated signals are then filtered with the same template bank that was used by the \texttt{pycbc} search algorithm during the second observing run of LIGO and Virgo, which is publicly available online~\citep{O2bankURL}. 
In the BNS region, the template bank uses the \texttt{TaylorF2} waveform model~\citep{Buonanno:2009zt}, and component masses in the range $1\msun<m<2\msun$, which results in a chirp mass range $0.87\msun<\mc<1.73\msun$. As mentioned above, spins are assumed to be aligned with the orbital angular momentum and in the range $-0.05\leq a \leq 0.05$. This parameter space is covered with $14,975$ templates.

For each simulated BNS, we use the \texttt{banksim} routine of the \texttt{pycbc} library~\citep{pycbc_code} to select the template in the bank that yields the best match, and we use the mass and spins of that template as the point-estimate that would be produced in low-latency (some caveats associated with our choices are listed in Sec.~\ref{Sec.Discussion}). 
The starting frequency of the overlap integrals needed to calculate the matched filter SNR is 27~Hz, following the convention used in the construction of the bank~\citep{pycbc_template_bank}. Simulated detector noise is not added to the signals before computing the overlap.

\begin{figure}[htb]
    \centering
    \includegraphics[width=0.5\textwidth]{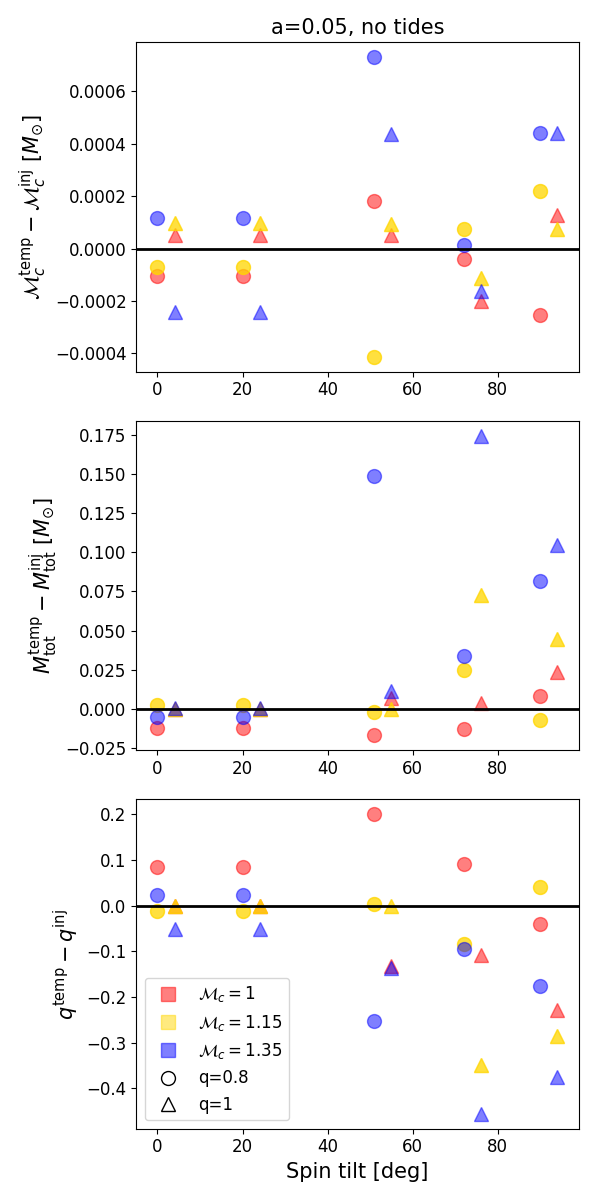}
    \caption{Difference between the value of parameters measured by the template bank and the true value, as a function of the spin tilt angle. An artificial offset in tilt angle is introduced for the equal-mass sources for clarity.
    The simulated BNSs have no tides and dimensionless spin $a=0.05$. 
    The values of the true chirp mass and mass ratio are given in the legend. 
    These results can be seen as the bias due to missing physics in the template bank.}
    \label{fig:q1_no_tides_bs}
\end{figure}

In the top panel of Fig.~\ref{fig:q1_no_tides_bs} we show the difference between the point-estimate and the true value of the chirp mass, plotted against the true value of the spins tilt angle for the simulated BNSs.
Triangles refer to equal-mass BNSs, whereas circles are used for the $q=0.8$ sources. 
Different colors refer to the true value of the chirp mass.

We see that no matter the true value of the chirp mass, mass ratio, and spin orientation, the difference between the measured and the true chirp mass is smaller than $10^{-3}~\msun$, and usually smaller than $5\times10^{-4}~\msun$.
The offsets are larger for the total mass (Fig.~\ref{fig:q1_no_tides_bs} middle panel) though they are still smaller than 6\% of the true value for all sources.

\begin{figure}[htb]
    \centering
    \includegraphics[width=0.5\textwidth]{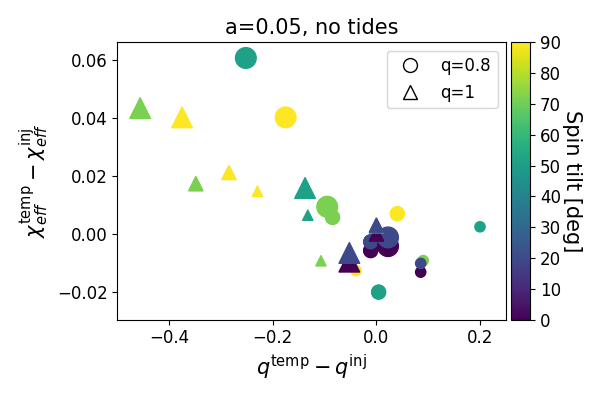}
    \caption{Correlation between the mass ratio and \chieff biases. The true
    value of the spin  tilt angle is given in the color bar. The size of the
    markers is representative of the chirp mass, from $\mc=1~\msun$ for the
    smallest markers to $\mc=1.35~\msun$ for the largest ones. The simulated
    BNSs have no tides and dimensionless spin $a=0.05$.}
    \label{fig:chi_q_corr}
\end{figure}

Finally, in the bottom panel of Fig.~\ref{fig:q1_no_tides_bs} we show the offset for the mass ratio $q$. 
For this parameter larger biases are visible when the spin tilt angles are larger than $20^\circ$. 
This is not unexpected since the mass ratio and the spins are correlated in the inspiral phase of CBC signals~(e.g. \cite{mass_ratio_spin1, mass_ratio_spin3, mass_ratio_spin2, 2013PhRvD..87b4035B,mass_ratio_spin4, 2018PhRvD..98h3007N}).
The biases in $q$ will usually be correlated with similar offsets in \chieff,
the mass weighted component of the total spin projected along the angular
momentum~\citep{2008PhRvD..78d4021R}. As shown in Fig.~\ref{fig:chi_q_corr}, when the spin tilt angles are larger than a few tens of degrees, significant biases appear in both the mass ratio and \chieff.

Next we modify our simulated signals to include tidal effects in the neutron stars. 
For all sources, we use the APR4~\citep{APR4} equation of state, which results in a dimensionless tidal deformability, $\Lambda$,~\citep{ PhysRevD.77.021502, PhysRevD.81.123016, PhysRevD.83.084051, lambda} in the range $[125,809]$, depending on the neutron star masses. 
We do not find that the results are significantly different than what we obtained in the absence of tides in the simulated signals, which is not surprising since tidal effects modify the phase evolution in the late inspiral (see e.g.~\cite{tidal_effects1, tidal_effects4, tidal_effects3, tidal_effects2}), at frequencies where LIGO and Virgo are not very sensitive. 
Given that the results are basically identical, we do not show plots for this setting.

\begin{figure}[htb]
    \centering
    \includegraphics[width=0.5\textwidth]{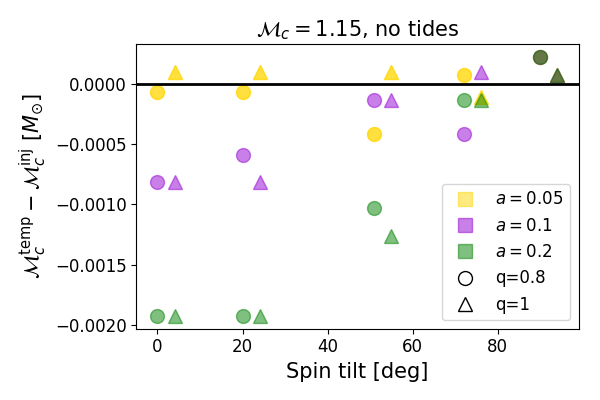}
    \caption{Same as in Fig.~\ref{fig:q1_no_tides_bs}, for BNS sources with spin magnitudes from $0.05$ to $0.2$. 
    The template bank only allows for spins up to $0.05$. 
    These simulated signals have $\mc=1.15~\msun$ and no tidal effects. 
    The values of the true dimensionless spin and mass ratio are given in the legend.}
    \label{fig:q0.8_bigger_spin_bs}
\end{figure}

We then proceed to verify the effect of neutron star dimensionless spins larger than the maximum value allowed in the template bank, i.e. 0.05. 
In fact, while $0.05$ is a value informed by the maximum spin of observed galactic double neutron star systems, pulsars have been observed with larger spins, the exact maximum depending on the (unknown) equation of state~\citep{pulsar_spins, bns_spins}. For these runs we only analyze sources with $\mc=1.15~M_{\odot}$.
In Fig.~\ref{fig:q0.8_bigger_spin_bs} we show the offsets for the chirp mass for systems with true spin magnitudes of $0.05$ (gold), $0.1$ (purple) and $0.2$ (green), plotted against the true value of the spin tilt angle,for both mass ratios $q=0.8$ (circles) and $q=1$ (triangles).
We stress again that the template bank does \emph{not} extend to spins larger than 0.05 in the BNS region, although the best match could be provided by a template in the NSBH region with larger spins for the same chirp mass but more unequal mass ratios. 
For these runs there thus are two different sources of potential bias: the spin magnitudes are too large for the template bank and the spin vectors can be misaligned. 
The results reported in Fig.~\ref{fig:q0.8_bigger_spin_bs} might be surprising at first, since they show that the chirp mass bias \emph{decreases} as the tilt angles of the spins go to $90^\circ$, i.e. to a configuration that yields the largest amount of orbital precession, which is a feature missing from the template bank.
These results can be interpreted by remembering that \chieff is the spin parameter that is best measured with GW data. 
Conversely, the component of the spins perpendicular to the angular momentum is usually poorly measured~\citep{salvo_spin,Farr_spin,purrer_spin, gwtc1,gwtc1_bbhPop}.
\chieff modifies the duration of the signal (keeping everything else the same,
larger and positive \chieff yields longer GWs as the system radiates less
efficiently~\citep{Campanelli}) and is conserved up to the second post-Newtonian order~\citep{2008PhRvD..78d4021R, 2014LRR....17....2B}. 
Since it affects the waveform evolution so much, \chieff is usually well measured~\citep{2017PhRvD..95f4053V,gwtc1,gwtc1_bbhPop, 2018PhRvD..98h3007N}.
On the other hand, the effects of spin precession in the detector frame are suppressed in systems for which: the spin magnitude is small; the orbital inclination angle is not close to $90^\circ$; the mass ratio is not very different from 1. Neutron star binaries suffer from all these limitations.

The results in Fig.~\ref{fig:q0.8_bigger_spin_bs} can thus be explained as follows. 
When the spin vectors are aligned with the angular momentum (i.e. the spin tilts are 0) then the entirety of the spin vectors contribute to \chieff, resulting in a value that cannot be matched by the template bank, unless $a=0.05$.
Physically, the source signal is longer than any waveform in the template bank with the same masses, due to its \chieff. 
To compensate for the missing waveform length,  a template with a smaller chirp mass is chosen, since these also yield longer signals~\citep{Maggiore}. 
This explains why for small tilt angles the template bank yields a point estimate of the chirp mass that is smaller than the true value.
As the true value of the tilt angle increases, the spin vectors contribute less and less to \chieff, leaving the waveform duration and phase evolution (at the 2nd post-Netwonian order) unaffected and removing the need to bias the chirp mass. While the trends in Fig.~\ref{fig:q0.8_bigger_spin_bs} are interesting, the  main conclusion remains the same: no matter the value of the tilt angle, the chirp mass is not biased to a level larger than $2\times 10^{-3}~\Msun$.
For the total mass offset, we obtain results which are similar to what was shown in Fig.~\ref{fig:q1_no_tides_bs} for sources with $\mc=1.15~\msun$.
We find that the mass ratio is overestimated for small tilts for the BNSs with $a=0.1$ and $a=0.2$. While the chirp mass bias can modify the waveform duration, the part of \chieff that cannot be matched by the template bank will also create a dephasing. At the lowest order in the post-Netwonian phase, that can be at least partially compensated for by overestimating the mass ratio (e.g. \cite{2013PhRvD..87b4035B} and \cite{2018PhRvD..98h3007N} eq. A2).


To assess the possible impact of the choice of the waveform model on the results we have presented, we have repeated a {subset of the tests described above using the \texttt{TaylorF2} waveform model for both the simulated BNS signals and the template bank. Specifically, we re-analyzed all combinations of the $q=0.8$ sources and the $q=1$ sources with $a=0.05$, no tides, and aligned spin (due to the limitation of the waveform) and find that this does not result in any significant difference with respect to the results we have presented above.}

\subsection{Results from full source characterization}

The results we have presented in the previous section answer the following question: how much bias can there be in the point estimate provided by CBC searches based on current template banks?
We have found numbers that are small for the chirp mass and the total mass, and potentially larger for the mass ratio.

We now want to address a different question: how do the biases we found compare with the \emph{statistical} uncertainty on the measurement of the same parameters that one would obtain using the full posterior distribution from higher-latency parameter estimation ({PE}) codes with waveforms that include all the relevant physics~\footnote{Most waveform models available for general use implement significant simplifications in their description of the physical processes governing the binary evolution, especially regarding the detailed microphysics of the compact objects near merger. See~\cite{Samajdar:2018dcx, Samajdar:2019ulq} for a discussion of BNS waveform systematics, and ~\cite{Shibata:2017xdx, Dietrich:2017xqb, Dietrich:2018phi, Duez:2018jaf} for a current overview of the status of BNS numerical relativity simulations.}?

To answer this question we have run the parameter estimation algorithm \texttt{bilby}~\citep{bilby} using the \texttt{dynesty} sampler~\citep{dynesty} and produced posterior distributions for all the extrinsic and intrinsic parameters on which BNS systems depend. 
These simulations are carried out by using the reduced order quadrature likelihood (ROQ)~\citep{roq} for the \texttt{IMRPhenomPv2\_NRTidal} waveform family for \emph{both} the source signal and the template signal used in the parameter estimation algorithm. We do not add simulated detector noise to the signals to be consistent with the \texttt{banksim} calculation, and start the integration at 32~Hz, which is the lowest frequency allowed by the ROQ likelihood~\footnote{If the parameter estimation analysis started at lower frequency, the statistical uncertainties would decrease, owing to the larger number of waveform cycles that can matched. In the context of our study, that would only result in the biases found in Sec.~\ref{sec.ResultBanksim} to be statistically more significant.}.
{As we have seen above, the choice of the waveform family is not a significant source of systematic errors for the BNSs we are working with. We thus do not rerun all of the PE analyses with different waveform models.}

To keep the computational cost contained, we follow the same strategy outlined above and run all combinations of mass ratio, chirp mass, and spin tilt angle only for the set of BNS for which $a=0.05$ and no tidal effects are added.
When adding tides or simulating BNS systems with spins of $0.1$ and $0.2$, we
only analyze sources with $q=0.8$ and the same combination of chirp masses as
used in the \texttt{banksim} analysis.

\begin{figure}[htb]
    \centering
    \includegraphics[width=0.5\textwidth]{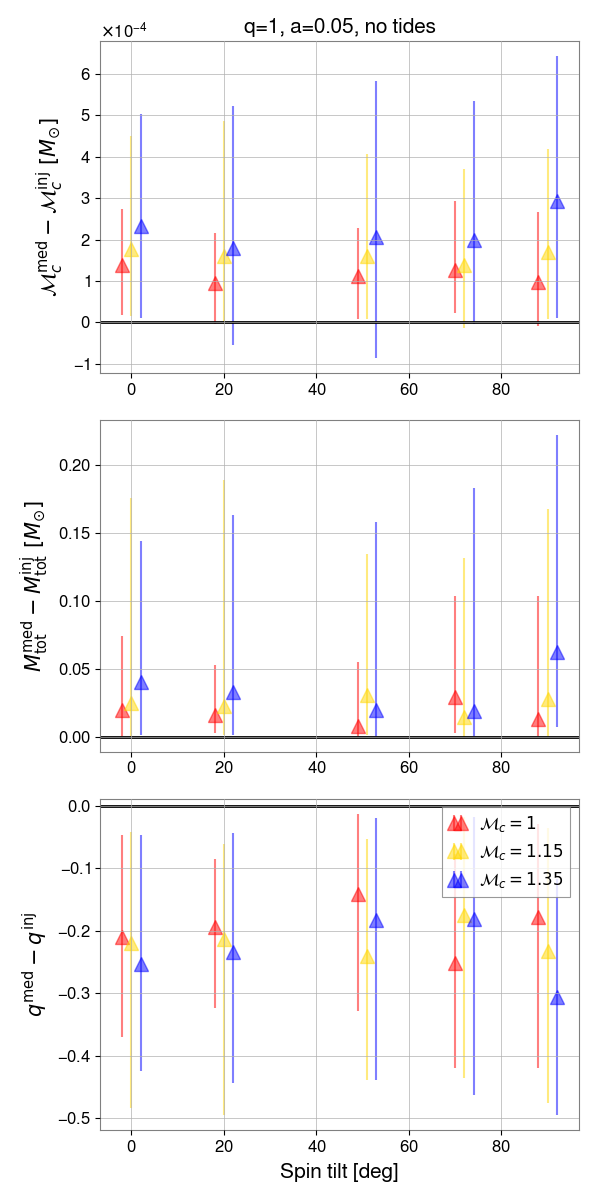}
    \caption{Difference between the median of the posterior from parameter estimation and the true value, as a function of the spin tilt angle. 
    The simulated BNS have no tides, $q=1$, and dimensionless spin $a=0.05$. 
    The error bars show the statistical 90\% credible intervals, while the marker reports the median. 
    The values of the true chirp mass is given in the legend.}
    \label{fig:q1_no_tides_pe}
\end{figure}

In Fig.~\ref{fig:q1_no_tides_pe} we report the median and the 90\% credible interval for the chirp mass (top panel), total mass (middle panel) and mass ratio (bottom panel) for the sources with $a=0.05$ and $q=1$, after subtracting the true values of the parameters.

We see that for the chirp mass estimate typical statistical uncertainties are of the order of $\mathrm{few}\times 10^{-4}~\msun$. 
Uncertainties are smaller for smaller chirp masses, as one would expect given that lighter objects have a longer inspiral, thus enabling a better measurement of their phasing. 
These uncertainties are of the same order of magnitude as the offsets from the template bank,  Fig.~\ref{fig:q1_no_tides_bs}.
In fact, for some of the sources the chirp mass point-estimate from the template bank lies outside of the 90\% credible interval from PE. 
This happens because the chirp mass is the parameter measured with the smallest statistical uncertainty for low-mass CBC signals, such as BNSs. 
While this implies that the biases on the chirp mass can be statistically significant, it must be stressed that one is still talking of offsets which are at most of the order of $10^{-3}~\msun$. 
This bares no practical consequences on the strategies for EM follow-up proposed by~\cite{M5}, which are based on assigning BNS sources to regions in the chirp mass space with characteristic widths of $\sim 0.08~\Msun$ or larger. 
It is clear that biases smaller than $\sim 10^{-3}~\msun$ will not move sources from one region to the other.

Next, we look at the mass ratio $q$, Fig.~\ref{fig:q1_no_tides_pe} bottom panel. 
We find that typical uncertainties are of the order of $\sim 0.4$ which is comparable to the size of the offsets we have found for equal-mass sources in the previous section. 
Finally, we look at the total mass, for which the statistical uncertainties are of the order of $\sim 0.1~\Msun$ or larger. 
Comparing this with Fig.~\ref{fig:q1_no_tides_bs}  we see that the offsets on total mass are usually smaller than the statistical uncertainty from PE. 

We find similar results for the systems with [$q=0.8$, $\mc=(1,1.15,1.35)~\msun$, $a=0.05$] with or without tidal terms, and for the systems with [$\mc=1.15~\msun$, $q=0.8$, $a=0.1,0.2$] for which tidal terms were not added.


\section{Discussion}\label{Sec.Discussion}

It has been recently argued that an estimation of the chirp mass for BNS mergers detected in gravitational-wave data could help prioritize their EM follow-up, if released in low-latency, together with a sky map and a distance estimation~\citep{M5}.

Two possible concerns with releasing chirp mass estimates produced in low-latency is that these are point-estimates calculated from template banks with finite resolution and that the waveforms used for the bank have missing physics; namely they do not account for neutron star spins larger than 0.05, spin precession, and tidal effects.
Theoretically, this all can lead to biases in the estimation of the source parameters: that is, the template in the bank that best matches the source signal might have parameters which are different from those of the source.

In this paper we have built a set of simulated BNSs with chirp masses in the range $[1-1.35]~\msun$, mass ratios in the range $[0.8-1]$, spin magnitudes in the range $[0.05-0.2]$ and spin tilt angles in the range $[0^\circ-90^\circ]$ and we have used the same template bank used by the \texttt{pycbc} search algorithm during the LIGO-Virgo second observing run (O2) to quantify eventual biases. 
All of the simulated events had a network SNR of 35 and an orbital inclination angle of $35^\circ$. 
The simulations were performed using \texttt{IMRPhenomPv2\_NRTidal} waveforms.

The O2 template bank consisted of $14,975$ \texttt{TaylorF2} waveforms in the BNS mass range, with spins exactly aligned with the orbital angular momentum and with magnitude $a\leq 0.05$, and no tidal deformability. 

We found that the bias on the chirp mass, i.e. the difference between the
point-estimate from the template bank and the true value, is always smaller
than a $\mathrm{few}\times 10^{-3}~\msun$ (and usually much smaller than that).

We have also shown that the eventual presence of tidal effects in the true signal and the exact waveform model used do not modify these conclusions.
The total mass of the BNS sources we simulated is also measured with a discrete accuracy by the template bank, with biases smaller than 6\% in all cases.
Conversely, the mass ratio and the effective inspiral spin, \chieff, can be more affected by missing physics in the template bank and can suffer from large biases.

We have compared the biases obtained from the template bank with the statistical uncertainties from a stochastic sampler~\citep{dynesty} which uses the same \texttt{IMRPhenomPv2\_NRTidal} waveforms used to simulate the  BNSs. 
We find that for the total mass, mass ratio, and chirp mass, the biases from the template bank are comparable to the size of the 90\% credible intervals from PE. 
However, for the chirp mass the biases are small enough that they would not have any practical implication on the EM observing strategies outlined in~\cite{M5}. 
The same is probably true for the total mass~\citep{metzger_private}, for which we get biases smaller than 6\%.

We conclude by mentioning a few caveats with our approach and results. 
We have assumed that the template which returns the largest overlap (i.e. recovers the larger fraction of SNR) will be selected by the search algorithm. 
In reality things are more complicated, and search algorithms go through a number of extra steps. 
For example, they require that the \emph{same} template is selected in all detectors in the network~\citep{mbta, pycbc_usman, spiir, gstlal_messick, pycbc_nitz, gstlal_sachdev}. 
Specific noise realizations at the individual detectors might thus result in selecting a different template than what would be chosen in a single-detector analysis.
While this is true, we notice that it is an effect that is not related to missing physics in the template bank, but rather to the properties of the noise, and that it would affect the PE step in a similar way, since this latter maximizes a joint likelihood across the detector network~\citep{lalinference}.
Thus, if the specific noise realization is a problem for the search, it will also be a problem for the PE step, and an eventual bias will not be reduced by waiting. 

Indeed, the chirp mass offsets we have found in this study are consistent with those of~\cite{Berry_2016}, where the \texttt{GstLAL} search algorithm was run on an astrophysical population of neutron star binaries (though covering a narrower parameter space than what we considered in this study). Those authors showed that their conclusions are virtually unchanged if one uses simulated Gaussian noise or real interferometric data.

A notable exception is if short instrumental noise artifacts exist in the stretch of data that contains the signal, which can be at least partially removed in higher latency, as was the case for GW170817~\citep{PankowMitigation}. 
In that case, depending on the shape of the artifact and how it overlaps with the inspiral signal, it might be the case that the search could give a biased result, while the PE step could give an unbiased one. 
Since in this paper we specifically focus on potential biases coming from missing physics, we did not expressly account for this possibility. 
However, this should become less and less of a concern in the future as low-latency methods to remove noise transients are being developed.

We have checked that the results we presented do not depend on the choice of the inclination angle for the simulated signals. 
Rerunning all template bank simulations with an inclination angle of $85^\circ$
yields virtually indistinguishable results, with the same template providing
the best match for all simulated signals as for the runs with an inclination
angle of $35^{\circ}.$ {Because the results do not change for this choice of
a nearly edge-on inclination where the effects of spin precession on the
waveform are maximal~\citep{mass_ratio_spin1,salvo_spin}, we do not expect variations at intermediate inclination angles.}

It should also be stressed that the search algorithms measure the detector-frame mass parameters, which are related to the source-frame masses (the astrophysically interesting values) by a factor of $(1+z)$, with $z$ the redshift of the source.
Even for GW170817, whose distance was only $\sim40$~Mpc, the difference between source-frame and detector-frame chirp mass was much larger than any of the biases we found in this paper~\citep{gw170817}. (This is not true for the total mass estimation.)
This is a difference that is easy to remove, as the LVC releases a marginalized distance posterior in low-latency that, once converted to a redshift posterior assuming a cosmology, can be used to transform detector-frame masses to source-frame masses. 
A related question is whether the statistical uncertainty on the estimation of the source-frame chirp mass, which includes contributions from both the detector-frame chirp mass and from the distance estimation, is so large that it makes it impossible to select a single bin of the ``multi-messenger matrix'' from \cite{M5}. We find that the statistical uncertainty on the source-frame chirp mass is on the order of $10^{-3}~M_{\odot}$, consistent with the offsets on the detector-frame chirp mass found in the low-latency search.

The results of this study suggest that the finite size of the template bank and the missing physics in the waveform models that are used do not introduce biases in the estimation of the chirp mass larger than $10^{-3}~\msun$. These are small enough that they cannot negatively impact the EM observing strategies outlined in~\cite{M5}.
It might be argued that even the total mass could be used to inform the EM follow-up strategies, as it doesn't suffer from biases larger than $6\%$. 


\acknowledgments

\textit{Acknowledgments.} S.B., S.V., and C.-J.H.~acknowledge support of the National Science Foundation, and the LIGO Laboratory. LIGO was constructed by the California Institute of Technology and Massachusetts Institute of Technology with funding from the National Science Foundation and operates under cooperative agreement PHY-1764464.
S.B. is also supported by the Paul and Daisy Soros Fellowship for New Americans and the NSF Graduate Research Fellowship under Grant No. DGE-1122374.
The authors would like to thank Tito Dal Canton, Ben Margalit, and Brian Metzger for useful comments and discussion. They also acknowledge Jolien Creighton, Heather Fong, and Thomas Dent for valued insights and Cody Messick and Christopher Berry for helpful comments on the manuscript.
The authors acknowledge the LIGO Data Grid clusters through NSF Grant PHY-1700765. 
LIGO Document Number P-1900239.

\bibliographystyle{apj}
\bibliography{chirp_mass}

\end{document}